\documentclass[12pt,preprint]{aastex}
\def\lea{\mathrel{<\kern-1.0em\lower0.9ex\hbox{$\sim$}}}
\def\gea{\mathrel{>\kern-1.0em\lower0.9ex\hbox{$\sim$}}}

\def\HST{{\it HST}}
\def\IUE{{\it IUE}}
\def\LS{LS~I +61~303}
\def\Swift{{\it Swift}}

\slugcomment{Submitted to The Astrophysical Journal}

\shorttitle{WeBo\,1 Companion}
\shortauthors{Siegel et al.}

\begin{document}

\title{Swift/UVOT Photometry of the Planetary Nebula WeBo\,1: Unmasking A Faint
Hot Companion Star}

\author{Michael H. Siegel\altaffilmark{1}, Erik Hoversten\altaffilmark{1},
Howard E. Bond\altaffilmark{2}, Michele Stark\altaffilmark{1,3}, and Alice A.
Breeveld\altaffilmark{4}}

\altaffiltext{1}{Pennsylvania State University, Department of Astronomy, 525
Davey Laboratory, University Park, PA 16802 (siegel@swift.psu.edu,
hoversten@swift.psu.edu)}
\altaffiltext{2}{Space Telescope Science Institute, 3700 San Martin Drive,
Baltimore, MD, 21218 (bond@stsci.edu)}
\altaffiltext{3}{Current Address: Department of Computer Science, Engineering,
\& Physics, University of Michigan-Flint, 213 Murchie Science Building, 303
Kearsley Street, Flint, MI 48502 (mistark@umflint.edu)}
\altaffiltext{4}{Mullard Space Science Laboratory/UCL, Holbury St.\ Mary,
Dorking, Surrey RH5 6NT, UK (a.breeveld@ucl.ac.uk).}

\begin{abstract}

We present an analysis of over 150 ks of data on the planetary nebula WeBo\,1
(PN G135.6+01.0) obtained with the \Swift\/ Ultraviolet Optical Telescope
(UVOT)\null.  The central object of this nebula has previously been described as
a late-type K giant barium star with a possible hot companion, most likely a young pre-white
dwarf. UVOT photometry shows that while the optical photometry is consistent
with a large cool object, the near-ultraviolet (UV) photometry shows far more UV flux
than could be produced by any late-type object.  Using model stellar atmospheres
and a comparison to UVOT photometry for the pre-white dwarf
PG~1159$-$035,  we find that the companion has a temperature of at least 40,000
K and a radius of, at most, 0.056 $R_{\sun}$.  While the temperature and radius
are consistent with a hot compact stellar remnant, they are lower and larger,
respectively, than expected for a typical young pre-white dwarf.  This likely
indicates a deficiency in the assumed UV extinction curve. We find that higher
temperatures more consistent with expectations for a pre-white dwarf can be
derived if  the foreground dust has a strong ``blue bump" at 2175~\AA\ and a
lower $R_V$. Our results demonstrate the ability of \Swift\/ to both uncover and
characterize hot hidden companion stars and to constrain the UV extinction
properties of foreground dust based solely on UVOT photometry.

\end{abstract}


\section{Introduction}
\label{s:intro}

The classical barium (or \ion{Ba}{2}) stars are red giants with enhanced
abundances of carbon and elements such as strontium and barium that are
synthesized in the $\it s$-process of neutron captures. First recognized by
Bidelman \& Keenan (1951), they are now understood as members of moderately wide
binary systems. When the more massive component became an
asymptotic-giant-branch (AGB) star, it dredged up carbon and $\it s$-process
elements to its surface and then transferred a portion of this material to the
companion through a stellar wind (e.g., McClure 1984; Jorissen et al.\ 1998;
Bond \& Sion 2001; and references therein). In this picture the AGB star has now
become an optically inconspicuous white dwarf (WD), leaving the optical light of
the system dominated by the contaminated cool companion. Strong support for this
picture came from the discovery by McClure (1984) that virtually all \ion{Ba}{2}
stars are single-lined spectroscopic binaries, with fairly long periods and
typical orbital separations of about 2~AU\null. 

In almost all cases, it is not possible to provide a direct confirmation of this
scenario by proving that the unseen companion star is a WD\null.  B\"ohm-Vitense
(1980) used the {\it International Ultraviolet Explorer\/} (\IUE\/) satellite to
detect a hot WD companion of $\zeta$~Cap, the prototypical \ion{Ba}{2} star.
Gray et al.\ (2011) recently showed that six barium dwarfs have UV excesses
consistent with the presence of a hot WD\null. However, in the large
majority of barium stars the WD---if that is what the companion is---has faded
below detectability, even in the UV, and we lack direct proof that the invisible
companions are really WDs. 

WeBo\,1 (PN G135.6+01.0; J2000: 02:40:14.4, +61:09:16) is a faint planetary
nebula (PN) that was discovered serendipitously by R.~Webbink (see Bond,
Pollacco, \& Webbink 2003; hereafter BPW03) during examination of Digitized Sky
Survey images of the X-ray binary LS~I +61~303 (V615~Cas). The X-ray source lies
only 4\farcm9 away from WeBo\,1. Deep narrow-band images of the PN have been
presented by BPW03 and Smith et al.\ (2007). The PN appears as a thin elliptical
ring, with a prominent 14th-mag central star. Spectroscopic observations by
BPW03 revealed that the nucleus is a cool barium star, making WeBo\,1 unique
among planetary-nebula nuclei (PNNi) known at the time of discovery. More
recently, however, a second \ion{Ba}{2} PNN, the central star of Abell~70, has
been discovered (Miszalski et al.\ 2012, who list several other cool PNNi
that may also have \ion{Ba}{2}-like compositions). In the picture of the
binary-star origin of \ion{Ba}{2} stars outlined above, we can argue that the PN
surrounding the barium star in WeBo\,1 must not only have been ejected during
the pollution process, with some of it accreting onto the optical star, but it
must be photoionized by the remnant of the AGB star, now a hot WD at the top of
the WD cooling track. Thus the cool star in WeBo\,1 would be an extremely young
\ion{Ba}{2} star, in the sense that the pollution of its surface must have
occurred very recently.

The only element missing in this seemingly satisfying story is direct proof that
there actually is a hot star present in the WeBo\,1 system. This cannot be shown
based on ground-based data, because the optical spectrum and colors of the
central star show only the cool barium giant. Ultraviolet (UV) observations from
space would be necessary to reveal the hot star. WeBo\,1 is at too low a
galactic latitude to be observed by {\it GALEX\/}, and has not been observed by
the {\it Hubble Space Telescope\/} (\HST)\null. However, there have been 
extensive observations of LS~I +61~303 and its surrounding field with the
\Swift\/ satellite's Ultraviolet Optical Telescope (UVOT)\null.  WeBo\,1 was
serendipitously within the UVOT field of view for these observations, and thus
there are many observations of it in the \Swift\/ archive.

In this paper, we present \Swift/UVOT photometry of WeBo\,1 in both optical and
near-ultraviolet (NUV) passbands, from which we confirm the presence of the
expected hot component. \S2 describes and presents the UVOT photometric data
analyzed in this paper. \S3 uses spectral models to reveal properties of the hot
companion star and examines the variability of WeBo~1. \S4 then summarizes our results.  We will use ``WeBo\,1'' to
designate both the PN and its central star, but the latter has also been
cataloged as V1169~Cas.

\section{Observations and Data}
\label{s:obsred}

\begin{figure}
\epsscale{1}
\plotone{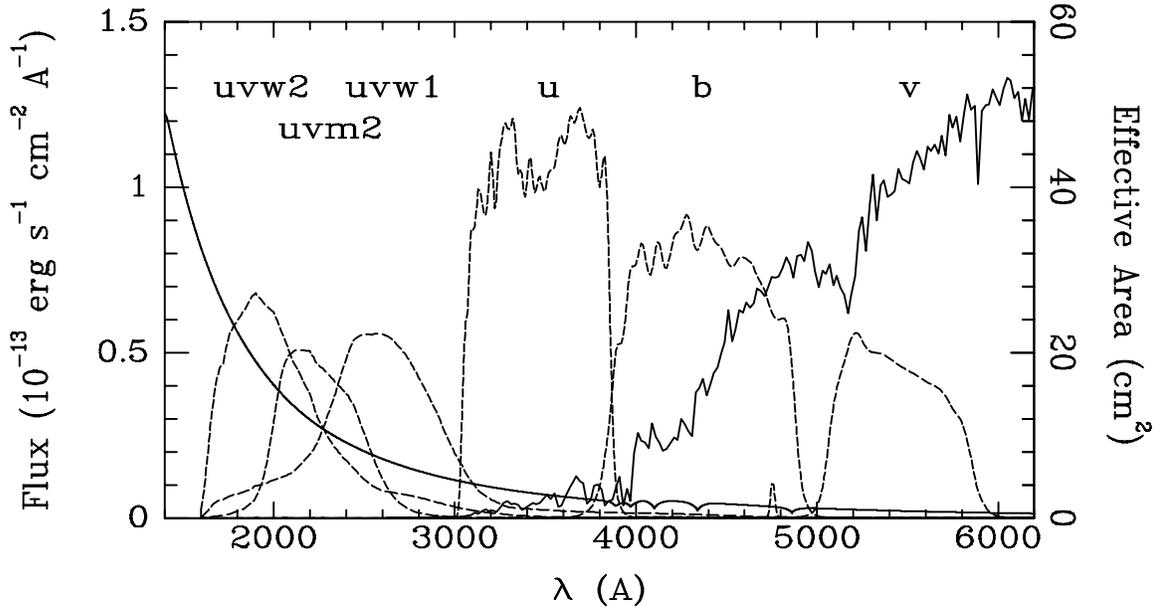}
\caption{A comparison of the spectra of a hot white-dwarf (WD) star to that of a
K~giant in the bandpasses of the \Swift\/ UVOT camera.  The WD spectrum (solid
line on left) is a model of the DA-type UV standard star
SDSS~J150050.71+040430.0 (WD~1458+042) from Siegel et al.\ (2010).  The model
spectrum is from the TLUSTY code (Lanz \& Hubeny 1995), with parameters set to 
$T_{\rm eff}=23,300$~K, $\log g=7.8$.  The K-giant spectrum (solid line on
right) is taken from the atmospheric models of Castelli \& Kurucz (2003), using
$T_{\rm eff}=4250$~K and $\log g=2$. It has been arbitrarily scaled to match the
flux of the WD at about 3800~\AA\null.  The dashed lines are the effective areas
of the six \Swift\slash UVOT filters used in this study.}
\label{f:spec}
\end{figure}

Our analysis of WeBo\,1 is based on data taken with the UVOT instrument aboard
the {\it \Swift\/} Gamma Ray Burst Mission (Gehrels et al.\ 2004).  UVOT is a
modified Ritchey-Chretien 30~cm telescope that has a wide ($17' \times 17'$)
field of view, which is imaged by a microchannel-plate intensified CCD operating
in photon-counting mode (Roming et al.\ 2000, 2004, 2005). The camera is
equipped with a filter wheel that includes a clear white filter, $u$, $b$, and
$v$ optical filters, $uvw1$, $uvm2$, and $uvw2$ UV filters, a
magnifier, two grisms, and a blocked filter.   Figure \ref{f:spec} illustrates
the bandpasses of UVOT's filters, and shows that they are well-positioned to
separate the UV and optical fluxes from two stars of significantly different
temperatures.

Although UVOT's primary mission is to measure the optical\slash UV afterglows of
gamma-ray bursts, the wide field, 2\farcs3 resolution, broad wavelength range
(1700-8000~\AA), and ability to observe simultaneously with \Swift's X-Ray
Telescope (XRT; Burrows et al.\ 2005) allow a broad range of investigations,
including the study of hot or energetic stars.  Our recent catalog of UV
photometric standard stars (Siegel et al.\ 2010) demonstrates the ability of
UVOT to measure the properties of hot compact stars, such as the one suspected
to be a companion to WeBo\,1.

WeBo\,1 has never been deliberately targeted by \Swift\/.  However, as noted
above, the nearby high-mass X-ray binary LS~I +61~303 {\it has\/} been monitored
extensively, and WeBo\,1 is well within the $17'$ UVOT field of view. Our search
of the \Swift\/ archive yielded a total of 160~ks of full-frame UVOT exposure
time on WeBo\,1, taken between 2007 November~1 and 2012 February~3. A large
fraction of these observations was performed using a UV-weighted six-filter
mode, while others were performed with UVOT in ``Filter of the Day"---a mode in
which only one of the $u$, $uvw1$, $uvm2$, or $uvw2$ filters is used, both to
save on filter-wheel rotations and to slowly build a UV sky survey.  The result
is that the accumulated exposure time on WeBo\,1 is heavily weighted toward the
UV filters.

\begin{deluxetable}{cccc}
\tablewidth{0 pt}
\tablecaption{\Swift/UVOT Observations of WeBo\,1\label{t:swiftphot}}
\tablehead{
\colhead{Filter} &
\colhead{Exposure Time (ks)} &
\colhead{AB Mag} &
\colhead{Var}}
\startdata
$v$    &  8.8 & $14.52\pm0.02\pm0.01$ &  1.11\\  
$b$    &  9.2 & $16.03\pm0.02\pm0.02$ &  1.14\\  
$u$    & 22.4 & $18.09\pm0.02\pm0.02$ &  1.14\\  
$uvw1$ & 28.3 & $19.62\pm0.02\pm0.03$ &  1.05\\  
$uvm2$ & 44.0 & $21.18\pm0.03\pm0.03$ &  1.17\\  
$uvw2$ & 47.2 & $20.24\pm0.02\pm0.03$ &  1.14\\    
\enddata
\end{deluxetable}

Photometry for WeBo\,1 was measured from deep stacked UVOT images using the
UVOTSOURCE program.  UVOTSOURCE, an FTOOL released as part of the
HEASOFT \Swift\/ suite of software programs\footnote{HEASOFT software can be found at: http://heasarc.gsfc.nasa.gov/docs/software/lheasoft/}, 
measure aperture photometry and then corrects the count rates
for the flat-field correction, coincidence losses and sensitivity loss over time. It then
transforms the instrumental magnitudes to a standard AB- and Vega-magnitude
systems, using formulae
and calibration data from Poole et al.\ (2008) and Breeveld et al.\ (2010,
2011)\footnote{The most recent UVOT zero points are available online from the
UVOT Digest at {\tt
http://heasarc.gsfc.nasa.gov/docs/swift/analysis/uvot\_digest/zeropts.html}}. 
In Table~\ref{t:swiftphot} we list the total exposure times in each UVOT filter,
and the mean magnitudes for WeBo\,1 with both random and
systematic errors given.  The fourth column gives an index of
variability, which is the ratio of the observed photometric scatter to the
calculated formal error. Non-variables should have an index near 1.0, while
strongly variable stars will have an index of 3.0 or more.  The tight clustering
of these measures near 1.0 (with a mean of 1.13) indicates that we detected little
evidence for  variability of WeBo\,1 to within 0.03-0.05 magnitudes, a point we address further in
\S\ref{ss:variability}.

To check on potential X-ray emission from the binary system, we ran the
automated analysis pipeline of Evans et al.\ (2009) on the data from the
\Swift\/ XRT, which were taken simultaneously with the UVOT observations.  While
there are several X-ray sources in the field (most notably \LS\ itself), we do
not find any X-ray emission at the position of WeBo\,1.

The WeBo\,1 PN has an expansion age of about $12,000\pm6,000$~yr (BPW03). The
putative hot companion star would be expected to be a hot central star or a WD
near the top of the WD cooling sequence, with its exact location on its post-AGB
evolutionary track depending on the exact age of the star and its mass.  While
Siegel et al.\ (2010) demonstrated an ability to constrain the properties of hot
WD stars (10,000--30,000~K) from UVOT data, WeBo\,1 
is expected to be hotter than their calibration stars in order to ionize the surrounding PN
(see, e.g., Hugelmeyer et al.\ 2007).  A search of the
{\it Swift\/} archive identified only a handful of known WDs with both extreme
temperatures ($T_{eff} > 50,000$~K) and low reddening ($E(B-V < 0.1$) for
comparison.  The latter requirement is particularly critical, given the
uncertainties in the UV extinction curve discussed below.  We settled on
PG~1159$-$035 (GW~Virginis) as the best object with which to test our ability to
constrain the properties of extremely hot degenerate objects. PG-1159 is
hot (140,000~K, Jahn et al.\ 2007), well-studied, has published
spectra and, most important, has minimal foreground reddening.  PG~1159$-$035 is
a prototype of a rare class of very hot, hydrogen-deficient WDs and is known to
display non-radial pulsations with an amplitude of approximately 0.02~mag
(Winget 1991; Costa \& Kepler 2008).  PG~1159$-$035 is not surrounded by a PN,
but many of the PG~1159 class are known PNNi (e.g., Kohoutek 1-16, Grauer \&
Bond 1984;  RX~J2117.1+3412, Motch et al.\ 1993; PG 1520+525, Jacoby \& van de
Steene 1995; and several others). While the WeBo~1 core is unlikely to be, like PG~1159$-$035,
deficient in hydrogen, PG~1159$-$035 defines a probable
the upper limit on its potential
properties, with the WDs of Siegel et al.\ (2010) defining the lower
limit.

PG~1159$-$035 has
never been specifically targeted for \Swift\/ observations.  However, like
WeBo\,1, it was observed serendipitously by UVOT on 2005 December 12 and 2011
July 28, during observations of the fortuitously nearby active galaxy Mkn~1310.
Table \ref{t:PG1159phot} lists the photometry measured at both epochs of
PG~1159$-$035. We find no significant difference in the magnitudes at the two
epochs, as expected given the relatively low pulsation amplitude of PG~1159.

\begin{deluxetable}{ccccc}
\tablewidth{0 pt}
\tablecaption{\Swift/UVOT Photometry of PG~1159$-$035\label{t:PG1159phot}}
\tablehead{
\colhead{} &
\multicolumn{2}{c}{2005 December 12} &
\multicolumn{2}{c}{2011 July 28}\\
\colhead{Filter} &
\colhead{Exposure Time (ks)} &
\colhead{AB Mag} &
\colhead{Exposure Time (ks)} &
\colhead{AB Mag}
}
\startdata
$v$    & 1.29	 & $14.75\pm0.02\pm0.01$& 0.08 & $14.71\pm0.04\pm0.01$ \\ 
$b$    & 1.29	 & $14.26\pm0.02\pm0.02$& 0.08 & $14.26\pm0.03\pm0.02$ \\ 
$u$    & 1.29	 & $13.72\pm0.02\pm0.02$& 0.19 & $13.68\pm0.03\pm0.02$ \\ 
$uvw1$ & 2.46	 & $13.17\pm0.02\pm0.03$& 0.16 & $13.13\pm0.02\pm0.03$ \\ 
$uvm2$ & \nodata & \nodata              & 0.30 & $12.95\pm0.02\pm0.03$ \\ 
$uvw2$ & \nodata & \nodata              & 0.31 & $12.69\pm0.02\pm0.03$ \\ 
\enddata
\end{deluxetable}

\section{Analysis}
\label{ss:results}

\subsection{A UV-bright Companion}
\label{ss:companion}

BPW03 estimated a spectral type of K0~III:p Ba5 for WeBo\,1, indicative of a
cool K giant with C$_2$ absorption bands and a very strong line of \ion{Ba}{2} at
4554~\AA\null. From the spectral type and the observed color, they estimated an
intrinsic $(B-V)_0$ color of 1.15 and a foreground reddening of $E(B-V)=0.57$. 
Our broadband optical colors from \Swift\/ (Table~\ref{t:swiftphot}) are
consistent with the BPW03 measures.  

Based on the optical spectral type, we can use stellar-atmosphere models to
predict the UV flux of the K0 star. We used the ATLAS9 compilation of
Castelli \& Kurucz (2003). For our model star, we adopted
$\rm[Fe/H]=0$, $T_{\rm eff}=4750$~K, $\log g=2.0$, $v_{\rm turb} = 2 \,\rm km\,
s^{-1}$, and a mixing length of $1/H=1.25$. This model spectrum was used to
generate synthetic photometry, following the method described in Siegel et al.\
(2010) and using the AB-mag zero points of Breeveld et al.\ (2011). We then
estimated the foreground reddening by applying the Pei (1992) Milky Way
extinction curve to the model spectrum and forcing the predicted $b-v$ color to
match the observed value.  The $b-v$ index is our longest-wavelength \Swift\/
color and should have minimal contamination from the anticipated hot companion.
We deduced an intrinsic $b-v$ color of 1.10 and a foreground reddening of
$E(B-V)=0.70$, somewhat higher than obtained by BPW03.

Barium stars are known to have a broad absorption feature in their spectral
energy distributions, centered near 4000~\AA\ (Bond \& Neff 1969). The
Bond-Neff absorption lies partially within the $b$ filter (see
Figure~\ref{f:spec}) and it would be expected to be strong in WeBo\,1. This may
explain the higher inferred $E(B-V)$, compared to that estimated from the $B-V$
color, since the $B$ band has a longer effective wavelength than $b$. However,
our neglect of the Bond-Neff effect in the following discussion should have a
minimal impact.

While the main bandpasses of the NUV filters are in the near-UV, $uvw2$ and
$uvw1$ have significant red leaks at optical wavelengths.  These leaks have
been well characterized using \HST\/ spectrophotometry of cool late-type stars
(Breeveld et al.\ 2011). The red leak is not a significant problem for hot
stars,
but a cool and reddened K~giant would have such minimal intrinsic UV emission
that almost any detection in the NUV filter would be due to red leak.  To
indicate the scope of the problem, we used methods detailed in Brown et al.\
(2010)  to generate synthetic magnitudes for the K~giant using the entire filter
curve given in Breeveld  et al.\ (2011), and compared them to synthetic
magnitudes generated using filter curves truncated at 2500, 3000, and 3300 \AA\
for the $uvw2, uvm2$, and $uvw1$ filters, respectively.  We find that a cool
K giant would have so little intrinsic UV flux that, in the absence of a hot
companion, the red leak would  contribute 99\%, 65\%, and 90\% of the signal
detected in $uvw2$, $uvm2$, and $uvw1$.  By comparison, for a hot (40,000 K)
companion star, only 13\%, 1\%, and 8\% of its signal would come from the red
tail. Thus, a K~giant would be almost undetectable in the NUV filters
without a red leak, faint but detectable in all but $uvm2$ with the red leak, and
easily detectable in all filters with a hot companion star. These calculations
indicate that the properties of any complex system in the UVOT filters can be
understood only if careful attention is paid to the red-leak contribution, an
ability we demonstrated in Siegel  et al.\ (2010).

Table \ref{t:colcomp} compares the synthetic and observed NUV magnitudes of 
WeBo\,1  with model magnitudes normalized to match the $v$-band brightness.  The
comparison shows that WeBo\,1, in comparison to the model, has a strong flux
excess in all of the UV passbands, {\it well above the flux predicted from the
well-characterized red leak}.  This excess is particularly notable in the $uvm2$
filter, which has minimal red sensitivity and is therefore the filter
most sensitive
to the presence of a hot companion star. {\it The dramatic
excess of UV flux clearly indicates that WeBo\,1 has a hot companion.}

\begin{deluxetable}{ccc}
\tablewidth{0 pt}
\tablecaption{Comparison of Observed WeBo\,1 Magnitudes to a Synthetic K Giant 
\label{t:colcomp}}
\tablehead{
\colhead{Filter}&
\colhead{WeBo\,1 Observed} &
\colhead{Synthetic\tablenotemark{a}}}
\startdata
$v$     & 14.52 & 14.52	 \\
$b$     & 16.03 & 16.03	 \\
$u$     & 18.09 & 18.64	 \\
$uvw1$  & 19.62 & 20.40  \\
$uvm2$  & 21.18 & 25.64	 \\
$uvw2$  & 20.24 & 21.80	 \\
\enddata
\tablenotetext{
a}{\small{Model properties: $\rm [Fe/H]=0$, $T_{\rm eff}=4750$~K, $\log g=2.0$,
$v_{\rm turb} = 2 \,\rm km \, s^{-1}$, $1/H = 1.25$; reddened by $E(B-V)=0.70$
and normalized to the $v$ magnitude of WeBo\,1.}
}
\end{deluxetable}

Our measured properties for the WeBo\,1 primary are somewhat sensitive to the
assumed model parameters.   The fit reddening, in particular, varies from 0.3 to
1.1 magnitudes if the temperature is allow to vary from 4000~K to 5500~K\null. Is it
possible that a different type of single star could produce the observed optical
and NUV photometry, especially given the interplay with the red leak?

\begin{figure}
\epsscale{1}
\plotone{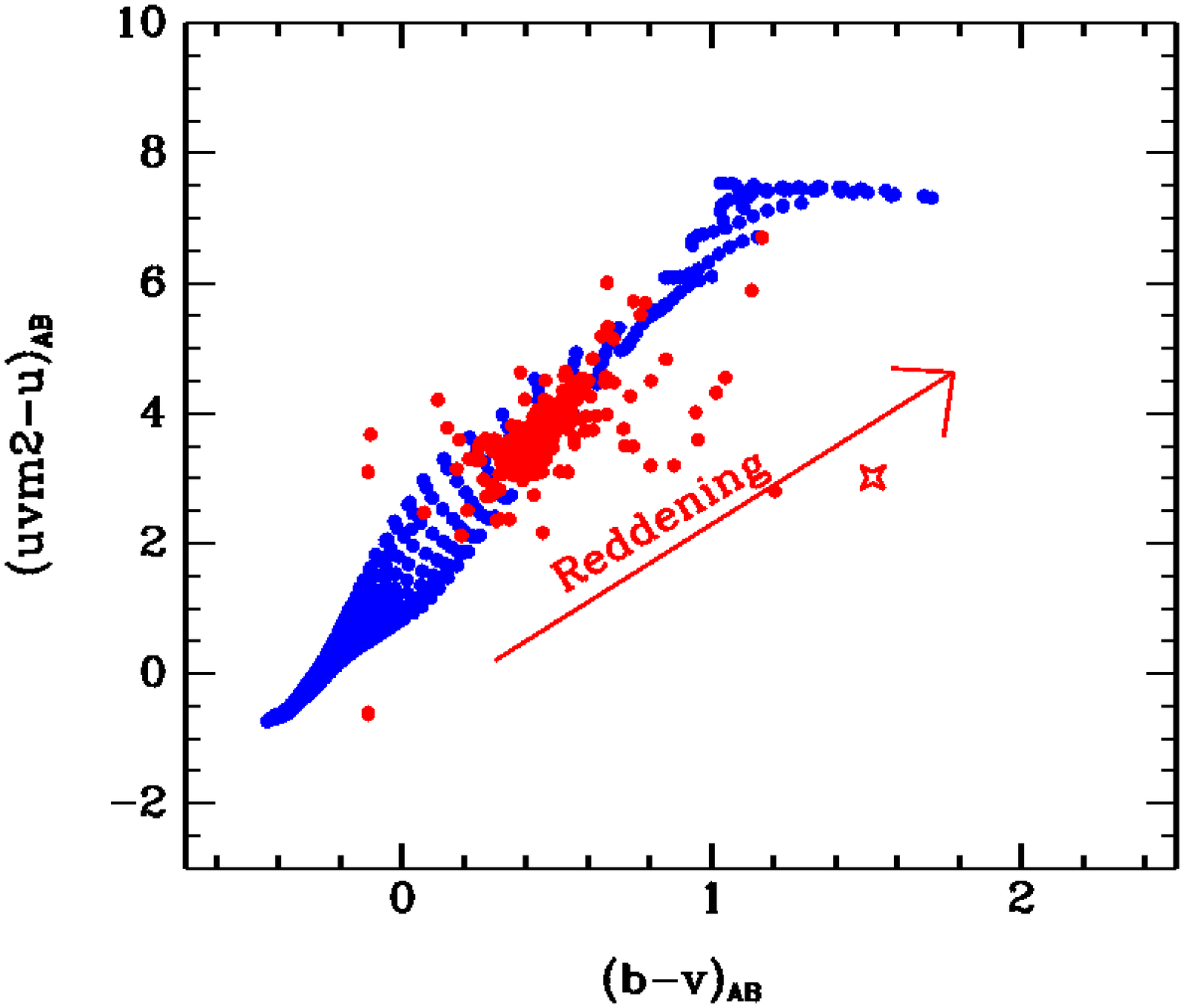}
\caption{
A comparison of the optical and NUV colors of WeBo\,1 to the intrinsic colors of
the Kurucz models of stellar atmospheres. The blue points represent the predicted
color for all of the available models with $\rm[Fe/H]=0$ and $v_{\rm turb}=2.0\,
\rm km\, s^{-1}$. The star is WeBo\,1.  Red points represent photometry of the
open cluster M67.  The line shows the reddening vector,
with the length of the line indicating the full Schlegel et al.\ (1998) extinction value along
this line of site.  Magnitudes are on the AB system.
}
\label{f:kzcomp}
\end{figure}

Figure \ref{f:kzcomp} shows the intrinsic (unreddened) $b-v$ and $u-uvm2$
colors of {\it all} of the Kurucz model atmospheres with $\rm[Fe/H]=+0.0$ and 
$v_{turb}=2.0\,\rm km \,s^{-1}$.  The models range from $\log g=0.0$, $T_{\rm
eff}=3500$~K to $\log g=5.0$, $T_{\rm eff}=50,000$~K\null.  Also shown for comparison is
UVOT photometry of the open cluster M67 (Siegel et al., in prep).  Note that with the filter curve properly modeled,
the photometric sequence in M67---an old
cluster of solar metallicity---matches the locus of synthetic magnitudes, lending confidence to our ability
to model the UVOT filters, including any red sensitivity.

No stellar model can reproduce both the optical and NUV colors of WeBo\,1
(starred point).  And no amount of assumed reddening can move WeBo\,1 onto the
Kurucz locus.  The arrow shows the reddening vector calculated from the Pei
models with the  length equal to the maximum reddening along the line of sight
from the Schlegel et al.\ (1998) maps ($E(B-V)=1.48$).  Only a significant and
large change to both the amount of foreground reddening {\it and\/} the
reddening law could possibly move WeBo\,1 even close to the locus of model
stars.  Varying the metallicity of the model family shifts the intrinsic color
locus only slightly and not nearly enough to capture WeBo\,1.  There is simply
no method by which the observed optical and NUV photometry can be fit by a
single star.  Two bodies---one cool and one faint but hot---are required.

\subsection{Comparison to PG~1159$-$035}

As discussed in \S2, we settled on PG~1159$-$035 as both a test of our 
ability to constrain the properties of extremely
hot compact objects and as a baseline against which to compare WeBo\,1.
However, confirming the nature of WeBo\,1 is more complicated than simply adding PG~1159$-$035 to the 
K giant, even assuming that such a comparison
were appropriate.  Shifting PG~1159's photometry to account for the
difference in reddening and distance does not reproduce the magnitudes and colors measured for WeBo\,1 (Table \ref{t:magcomp}).
This may be due to differences between the two objects but is even more likely due to the additional reddening.  Accounting for 0.70 magnitudes
of reddening in the NUV passbands is more complicated than simply adding
an offset to the photometry as we have done in Table \ref{t:magcomp}.  The NUV extinction curve is uncertain, particularly the
strength of the blue bump at 2175 \AA\null.  Moreover, the curve is steep and extinction can vary depending on spectral type.


\begin{deluxetable}{ccc}
\tablewidth{0 pt}
\tablecaption{Comparison of Observed WeBo\,1 to Synthetic K-Giant and 
Shifted PG~1159$-$035 Photometry\label{t:magcomp}}
\tablehead{
\colhead{Color}&
\colhead{WeBo\,1 Observed} &
\colhead{Synthetic\tablenotemark{a}}}
\startdata
$v$     &  14.52 & 14.52\\
$b$     &  16.03 & 15.94\\
$u$     &  18.09 & 17.83\\
$uvw1$  &  19.62 & 18.46\\
$uvm2$  &  21.18 & 20.18\\
$uvw2$  &  20.24 & 19.36\\
\enddata
\tablenotetext{a}{\small{Magnitude generated by adding a scaled model from Table \ref{t:colcomp} to PG~1159$-$035 
magnitudes from Table \ref{t:PG1159phot}. The latter were shifted by 1.4 magnitudes to account for difference
in distance and by UV extinction values for $E(B-V)=0.70$.}}
\end{deluxetable}

Properly accounting for all the potential confounding factors---red leak, UV
extinction curve, differences between the secondary star and
PG~1159$-$035---requires spectral synthesis, in which theoretical and
observational models are used to recreate the observed photometric measures, a
method we used with notable success in creating  WD UV standard stars (Siegel et
al.\ 2010) and which we used in \S3.1 to rule out a single star.

\subsection{Spectral Modeling of the Companion Star}

Revealing the origin of the UV excess in WeBo\,1 requires careful modeling of
both the cool giant, and the putative hot companion.  Modeling of the K giant is
informed by spectroscopic identification of the star.  As discussed earlier, the
K giant is modeled using a Kurucz model stellar atmosphere with $T_{\rm
eff}=4750$ K, $\log\ g=2.0$, and $v_{turb} = 2\,\rm km \,s^{-1}$.  This model
spectrum was used to generate synthetic magnitudes following the method
described in Siegel et al.\ (2010) and using the AB zeropoints of Breeveld et
al.\ (2011).

The UV excess suggests a companion to the K giant which is similar to the 
pre-WD PG~1159 star.  Unlike the K giant, the temperature of the companion is a
free parameter. Ideally, one would measure the properties of the companion by
comparison to a spectral library of extremely hot WD and pre-WD
stars, such as the one published from \IUE\ data (Holberg et al.\ 2003). 
However, these spectra only go as red as 3150 \AA, which does not cover all of
the UVOT filters and does not cover the red leak in the $uvw1$ and $uvw2$
filters.

\begin{figure}
\epsscale{0.8}
\plotone{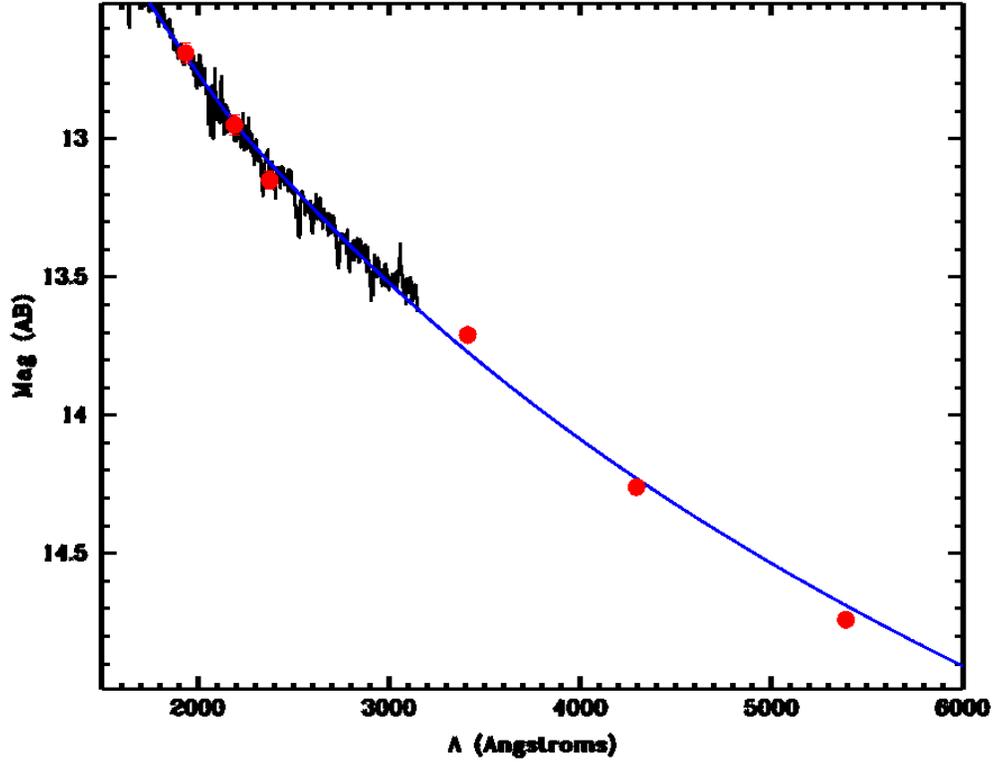}
\caption{Comparison of observed photometry and synthetic spectra for the pre-white dwarf
star PG~1159$-$035.  Solid red points are the photometric measures at the effective wavelengths for a 140,000 K
blackbody.  Point
sizes are comparable to the observational uncertainties.  The blue line is the blackbody curve for a 140,000 K blackbody
while the solid black line is the \IUE\/ spectrum of PG~1159$-$035 taken from 
Holberg et al.\ (2003).}
\label{f:PG1159comp}
\end{figure}

However, a simpler model may be sufficient for modelling the companion.  UV
spectra of hot compact  WDs and PG~1159 stars (Feibelman 1996; Kruk \& Werner
1998; Marcolino et al.\ 2007) indicate that the hottest stars should have few
absorption features in the UVOT wavelength range, which occupies the
Rayleigh-Jeans tail of the spectral energy distribution.  We ran two models of
PG~1159$-$035 through our spectral modeling software, one using a pure black
body of 140,000 K, the other using the published \IUE\/ spectrum over 
the 1100-3150
\AA\ range and a blackbody curve for the remainder.  In both cases, we
re-normalized the photometry to minimize $\chi^2$.  Figure \ref{f:PG1159comp}
shows the results.  Both the hybrid model and the pure blackbody model provide a
reasonable fit to the photometric measures, with reduced $\chi^2$ values of 3.2
and 3.5, respectively,  with the most significant discrepancy in the $u$-band
measure.  For simplicity's sake, we used blackbody models for the companion
star.

The K giant and hot companion models were fit to the observed photometry
simultaneously using the double-star $\chi^2$ method described in Hoversten et
al.\ (2008).  A standard $\chi^2$ fit to a model can be expanded to include two
models as follows:
\begin{equation}
\chi^2=\sum_i\left[\frac{F_{o,i}-af_{A,i}-bf_{B,i}}{\sigma_i}\right]^2 \, ,
\label{eq:multistar1}
\end{equation}
where $F_{o,i}$, $f_{A,i}$ and $f_{B,i}$ are the fluxes of the observed object, first template model and 
second template model in the $i$th bandpasss, $\sigma_i$ is the photometric uncertainty in the $i$th band, and $a$ and $b$ are the
flux weighting coefficients that minimize the $\chi^2$.  This equation can then be expressed in terms of magnitudes
\begin{equation}
\chi^2=1.086^2\sum_i\left[\frac{1-a10^{\left(m_{o,i}-m_{A,i}\right)/2.5}-
b10^{\left(m_{o,i}-m_{B,i}\right)/2.5}}{\sigma_{m_i}}\right]^2 \, ,
\label{eq:multistar2}
\end{equation}
where $m_{o,i}$ and $\sigma_{m_i}$ are the observed magnitude and magnitude error in the $i$th band, $m_{A,i}$, and $m_{B,i}$ are the model
magnitudes of the first and second models in the $i$th band, and $a$ and $b$ are again weighting coefficients that minimize $\chi^2$.

The relative weighting of the two models appears to be a free parameter, adding extra degrees of freedom to the $\chi^2$ minimization.  However,
this in not the case as $a$ and $b$ can be analytically determined by setting $\partial \chi^2/\partial a$ and $\partial \chi^2/\partial b$ equal to
zero and solving the simultaneous system of equations.  The result is that
\begin{equation}
a=\frac{AB^2 - BC}{A^2B^2 - C^2}
\label{eq:multistar3}
\end{equation}
and
\begin{equation}
b=\frac{BA^2 - AC}{A^2B^2 - C^2} \, ,
\label{eq:multistar4}
\end{equation}
where
\begin{equation}
A = \sum_i\left(\frac{10^{(m_{o,i}-m_{A,i})/2.5}}{\sigma_{m_i}^2}\right),
\label{eq:multistar5}
\end{equation}
\begin{equation}
B = \sum_i\left(\frac{10^{(m_{o,i}-m_{B,i})/2.5}}{\sigma_{m_i}^2}\right),
\label{eq:multistar6}
\end{equation}
and
\begin{equation}
C = \sum_i\left(\frac{10^{(2m_{o,i}-m_{A,i}-m_{B,i})/2.5}}{\sigma_{m_i}^2}\right).
\label{eq:multistar7}
\end{equation}
This shows that under the assumption of $\chi^2$ minimization there exists an optimal weighting of two spectral models.  Once $a$ and $b$ have been
calculated it is only necessary to consider one combination of any two models.

\begin{figure}
\plotone{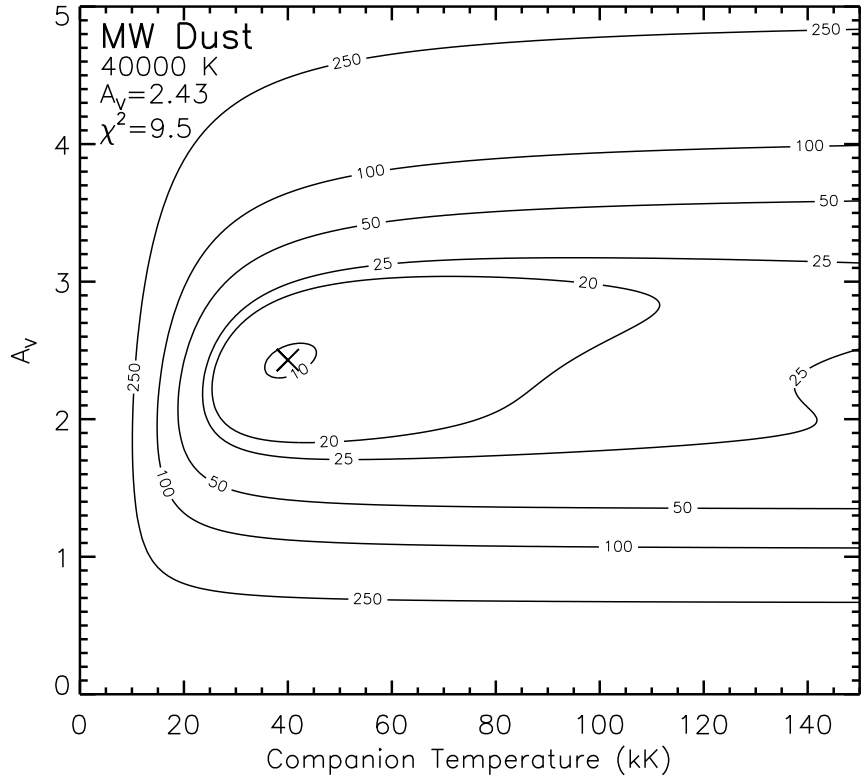}
\caption{Minimized $\chi^2$ values for WeBo\,1 assuming a K giant and a black
body companion as a function of temperature and foreground extinction.  This
plot uses Milky Way dust with a strong 2175 \AA\ bump.  Temperatures are in
units of 1000~K\null. The cross shows the best fitting companion model at 40,000
K and $A_V=2.43$.}
\label{fig:mwcontour}
\end{figure}

\begin{figure}
\plotone{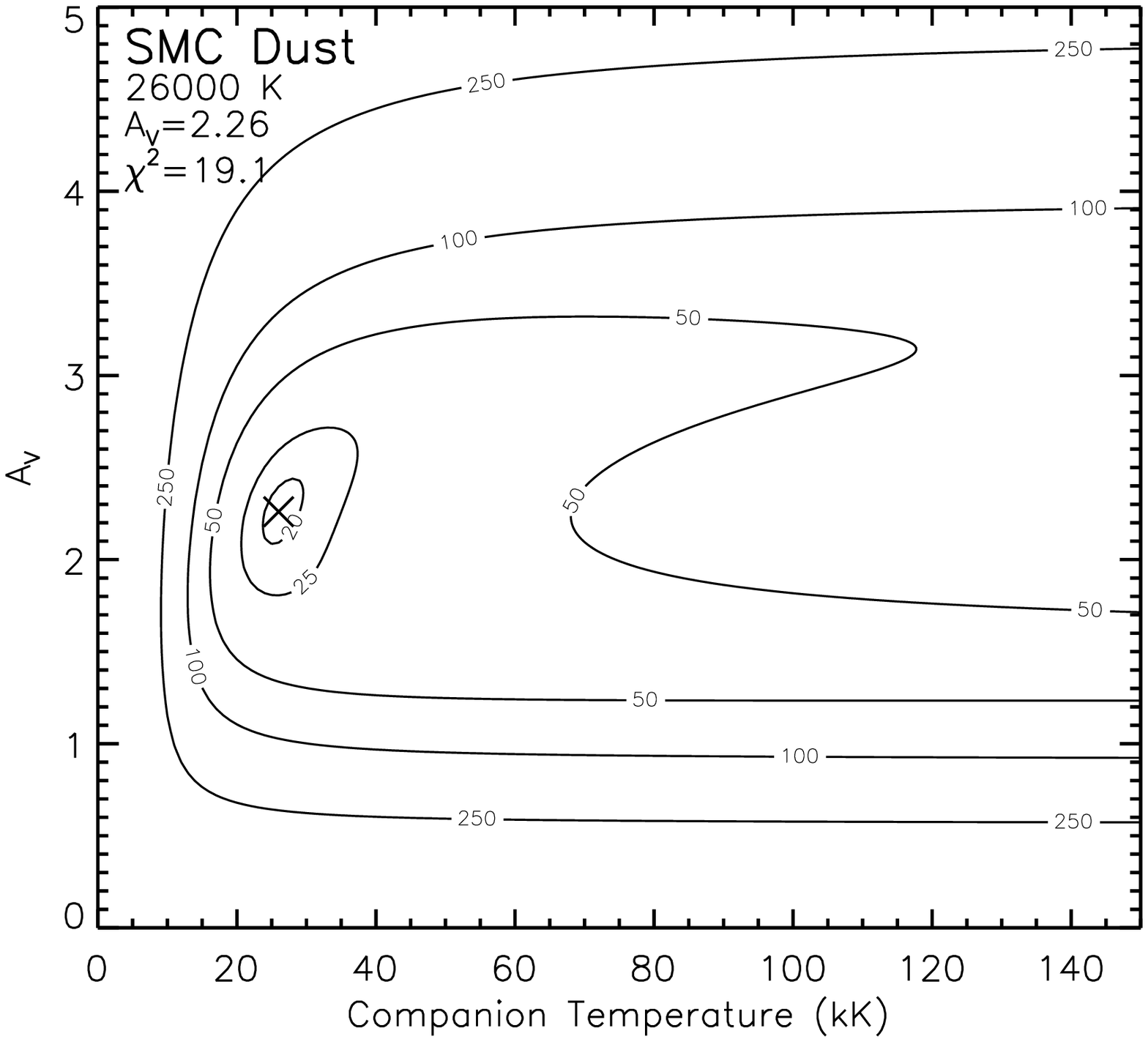}
\caption{Minimized $\chi^2$ values for WeBo\,1 assuming a K giant and a black
body companion as a function of temperature and foreground extinction.  This
plot uses SMC dust without a strong 2175~\AA\ bump.  Temperatures are in units
of 1000~K\null. The cross shows the best fitting companion model at 26,000 K and
$A_V=2.26$.}
\label{fig:smccontour}
\end{figure}

Figure \ref{fig:mwcontour} shows the $\chi^2$ space assuming Milky Way dust in
the foreground while Figure~\ref{fig:smccontour} shows a similar plot assuming
SMC dust (without a blue bump at 2175 \AA).  The models using Milky Way dust and
a strong blue bump are notably superior to those using an SMC dust model without
a blue bump.  For the former, the minimum $\chi^2$ is reached for a companion
temperature of 40,000 K and a foreground extinction of $A_V=2.43$.  This is
slightly higher than the extinction calculated just from the K giant itself
($A_V=2.17$).

\begin{figure}
\epsscale{0.8}
\plotone{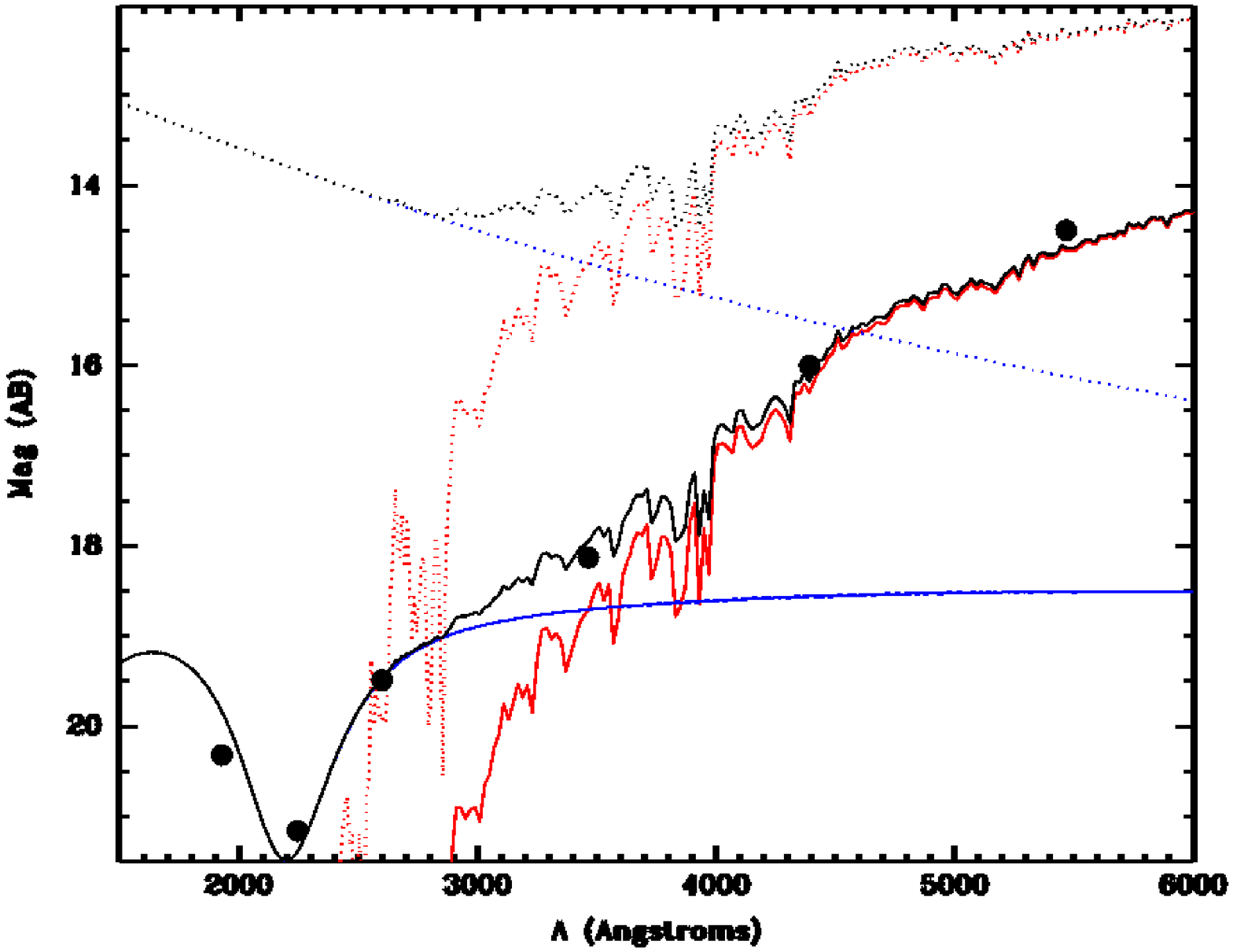}
\caption{Comparison of observed WeBo\,1 photometry to best fitting K giant plus
hot companion model.  Solid points are the data, with error bars smaller than
the size of the points.  Magnitudes are on the AB system. The lines are the
spectral models---blue for the hot companion, red for the  K giant and black
for the combined system.  Dotted lines denote the spectra in the absence of
foreground extinction.  Note the dramatic effect of the foreground dust,
particularly the 2175 \AA\ bump.}
\label{fig:magplot}
\end{figure}

Figure \ref{fig:magplot} compares the best fitting black body and K giant models
to the observed photometry of WeBo\,1, again illustrating that neither star can
provide an adequate representation of the measured photometry.  However, the
combined photometry matches the data points well.  We also show the unextincted
spectra to demonstrate just how dramatically the foreground dust
obscures the hot companion star.  Note the significant downturn
in the UV flux in the $uvm2$ passband; a result of the strong blue bump
in the foreground dust.  Note also that, under this model, we should see some emission
from the companion star near the H and K lines.  BPW03 did not see this and this could indicate
that WeBo's 1 companion is fainter and hotter than our fit models.

If the WeBo\,1 companion is a WD, having passed the turnaround point in the H-R
diagram, the 40,000 K temperature is problematic.  It would imply a low mass for
the companion star (below 0.3 $M_{\sun}$; Sch\"onberner et al.\ 1989; Panei et
al.\ 2007) and therefore a rather high  WD age (at least several 10$^7$ years). 
This would be incompatible with the 12,000 yr age of the nebula.  At 40,000 K, it
would be more likely that the WeBo\,1 companion is in a young PNN phase still
evolving to higher temperature and not having yet reached the turnaround point.

However, there are a number of reason to believe that the temperature of WeBo\,1 is higher than our nominal
estimate.  At this temperature, WeBo\,1 is at the limit
of UVOT's ability to constraint the properties of extremely hot
stars.  Note in Figures \ref{fig:mwcontour} and \ref{fig:smccontour} that the
$\chi^2$ contours are elongated in the temperature direction.  A star as
hot as WeBo\,1's companion is far brighter in the FUV than the NUV and we are
likely probing the Rayleigh-Jeans tail of the spectral energy distribution, which
has limited leverage on the temperature.  Running PG~1159$-$035 through the software
shows that we can establish a strong lower limit on the temperature ($\sim$ 100,000 K)
but no upper limit.  A small change in the underlying
models would not lower our estimated temperature significantly, but
could increase it dramatically.

It is worth noting that simply running a 140,000 K blackbody through
our engine provides a perfectly adequate fit to the photometry of PG~1159$-$035.
This indicates that we {\it should} be able to measure the temperature of
WeBo\,1 or at least place a better lower limit if it is indeed as hot as PG~1159$-$035. However, PG~1159$-$035 is a
simple system---a single star with minimal foreground reddening.  WeBo\,1,
by contrast, involves  an unusual barium star with a  high degree of
chromospheric activity (BPW03), a pre-WD star companion in an early stage
of evolution, and a high amount of foreground dust.  This system may simply be
too complex for the modelling software.

\begin{figure}
\plotone{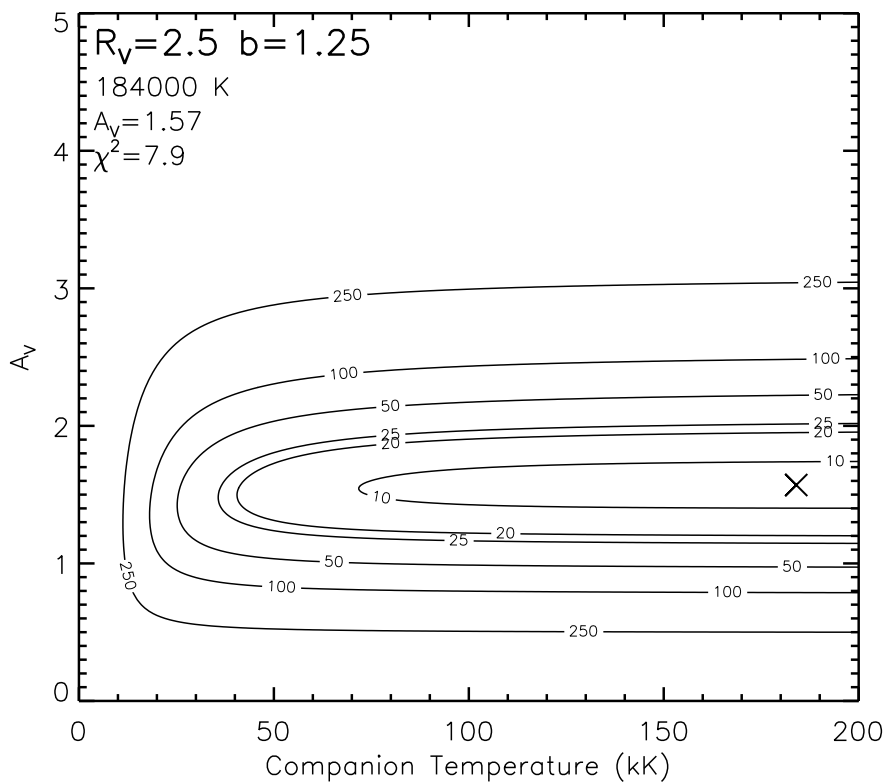}
\caption{Minimized $\chi^2$ values for WeBo\,1 assuming a K giant and a black
body companion as a function of temperature and foreground extinction.  This
plot uses Milky Way dust with a slightly stronger 2175 \AA\ bump and a lower
$R_V$, resulting in a steeper extinction in \Swift\/'s NUV passbands. 
Temperatures are units of 1000~K\null. The cross shows the best fitting
companion model at 184,000 K and $A_V=1.57$.  Not that the contours are narrowly
elongated in the temperature direction, indicating we are at or beyond the limit
of UVOT's sensitivity to stellar temperature.}
\label{fig:bumpcontour}
\end{figure}

The most likely culprit for this sabotage is the foreground dust.  The
extinction curve in the UV is still poorly known.  Simply adding in the blue
bump increases the estimated temperature of the companion by 14,000 K and cuts
the reduce $\chi^2$ in half.  Using techniques developed in our study of the
Milky Way dust (Hoversten et al., in prep), we varied the strength of the blue
bump and the assumed extinction law ($R_V$).  We find that for a blue bump
strength of 1.25 and a $R_V$ of 2.5, the derived temperature of the star shoots
up to 184,000 K and the foreground extinction is more consistent with that
derived from the K giant and by BPW03 (Figure \ref{fig:bumpcontour}).  These
parameters are well within the range of extinction laws described by Cardelli et
al.\ (1989).  The implication is that the inferred properties of such a hot star
are uniquely sensitive to the assumed properties of the foreground dust. If this
is the case, studies of more hot pre-WDs could provide powerful insight
into the Milky Way's extinction law in the UV.

At a temperature of 184,000 K, WeBo\,1's properties would be consistent with a
young WD or pre-WD\null. Its exact evolutionary
state would depend on its mass.

The normalization used for the spectral models intrinsically includes both the
radius and distance to the WeBo\,1 system.  Assuming a distance of 1600 pc
(BPW03), a Milky Way dust model, and calculating radii based on the $b$ and $v$
magnitudes for the  cool primary star and the $uvm2$ magnitudes for the
companion, we calculate stellar radii of 5.3 and 0.057 $R_{\sun}$,
respectively.  This is broadly consistent with expectations for a red giant and
WD star.  Applying this method to PG 1159$-$035 correctly recovers the
known stellar radius of 0.025 $R_{\sun}$.  If the alternative dust model is
assumed, the radii of the stars shrinks to 3.7 and 0.028 $R_{\sun}$ for the
K giant and pre-WD, respectively.

In sum, although our spectral models calculate a lower temperature
of the WeBo\,1 secondary than expected, the data are inconsistent with anything other than a
small, compact,  hot object as the companion to the red giant star.  However,
even a fairly small change in the assumed foreground dust results in a WeBo\,1
secondary that is similar in temperature and size to PG~1159$-$035.

\subsection{Photometric Stability}
\label{ss:variability}

BPW03 indicated that WeBo\,1 showed variability on a timescale of 4.7 days,
possibly as a result of starspots on a rotating primary star. As noted above, column four of
Table \ref{t:swiftphot} shows the ratio of observed to expected scatter.  In all
six passbands, WeBo\,1 shows minimal scatter beyond that expected from the
photometric errors.  However, this broad and somewhat inhomogeneous dataset may not be
ideal to detect the variability.  If the variability is indeed the result of starspots,
that signal may be blurred if the starspots changed over the five years. 

We investigated this further by calculating a Welch-Stetson (1993) variability
index for all two-filter combinations in our photometry. The Welch-Stetson index
measures the correlation of photometric residuals from the mean flux.  Pulsating
stars should have positive residuals or negative residuals simultaneously in
different passbands.  Non-variable stars should have an index near zero.  We found numerous
correlations among the filter combinations, ranging from 0.04 to 0.42.  Much of this correlation
is the result of a long-term fading of 1-2\% per year in all passbands except $b$ and $uvm2$.  It's
not clear what would cause this trend -- whether it is something secular connected with the WeBo\,1 primary
or something related to the calibration.  The decline in UVOT's sensitivity is well-characterized by Breeveld 
et al.\ (2011)
and incorporated into the current FTOOLS build.  However, a small error in sensitivity decline or large scale sensitivity
could potentially produce this signal.

\begin{figure}[ht!]
\includegraphics[scale=1.]{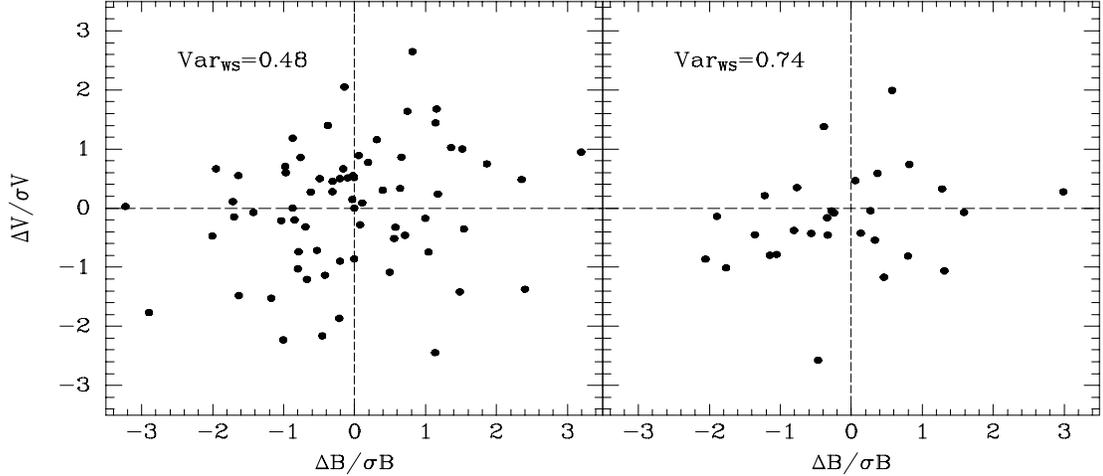}
\figcaption[fig8.eps]{Residual correlation in \Swift\slash UVOT $b$ and $v$
photometry of WeBo\,1.  The points represent $b$ and $v$ photometric measures
taken within 0.5 day of each other.  The axes measure the ratio of photometric
residuals to photometric uncertainties ($\Delta/\sigma$ where
$\Delta_V=V_i-\langle V\rangle$. For a truly variable star, the residuals should be
correlated resulted in a diagonal line from the lower left to upper right (see,
e.g., Figure 1 of Welch \& Stetson (1993). The residuals of WeBo\,1 show a
slight correlation with a positive variability index.  Narrowing the data to
those taken in late 2011 and early 2012 shows a stronger
correlation.\label{f:vbcor}}
\end{figure}

However, this long-term decline does not explain all of the variability.  The left panel of Figure \ref{f:vbcor} shows 
the residuals for the $b$ and $v$ passbands for the entire data set while the right panel shows those from 2011 October to 2012 February, when
{\it Swift\/} was monitoring LS~I +61~303 at least every week and, during 2012 December, every day.  For the latter time span,
we see a correlation between the residuals, which
hints at potential variability in the WeBo\,1 primary.  We attempted to fit this period using phase dispersion
minimization with the IRAF task PDM over periods between 0.1 and 100 days.  We found multiple solutions, none of which were particularly
superior or showed a particularly clear light curve.

It should be noted that the variability detected by BPW03 was of order 30 mmag.  This
is less than the random errors on individual observations of WeBo\,1, which have
a mean of 95, 46, and 41 mmag in the $ubv$ passbands, respectively.  Even
binning observations made on the same day reduces the mean error of the measures
only to 77, 37, and 32 mmag, barely sufficient to measure the variability. 
Furthermore, the \Swift\slash UVOT data are not optimized for the
detection of the primary star's variation.  There are gaps in the phase coverage
when \Swift\/ was either not observing V615 Cas or not observing it in the
optical filters that are the most sensitive to WeBo\,1's primary star.  The
modes it has been observed in are weighted heavily toward the NUV.  All of these factors
make the detection of WeBo\,1's variability difficult.

\begin{figure}[ht!]
\includegraphics[scale=1.]{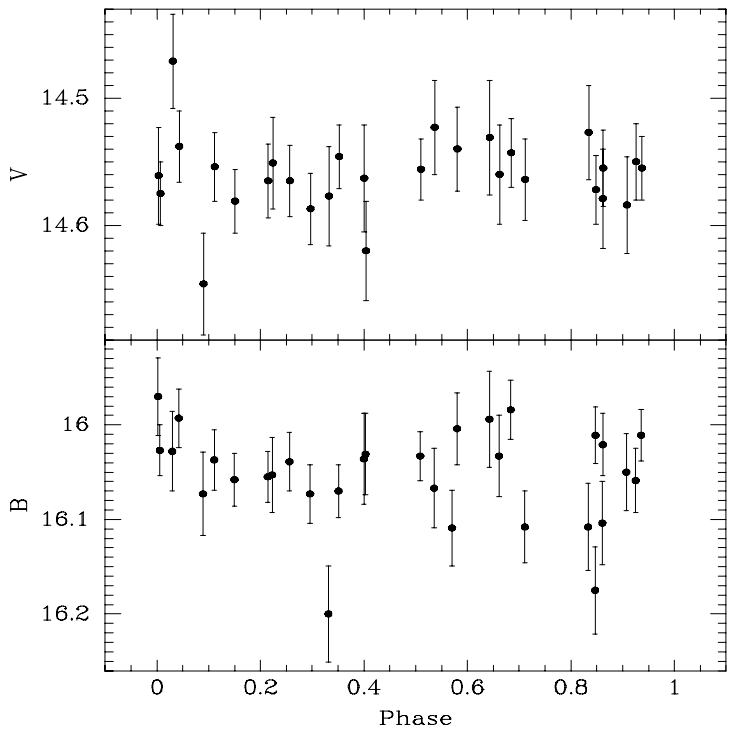}
\figcaption[fig9.eps]{$v$- and $b$-band photometry of WeBo\,1 folded with the
4.686 day periodicity detected by BPW03.  Photometric measures are binned by
calendar day.
\label{f:vv4686}}
\end{figure}

However, Figure \ref{f:vv4686} shows the $v$- and $b-$-band photometry from late 2011 and early 2012 folded with a period of
4.686~days, the periodicity identified by BPW03.  As can be seen, while a 30 mmag variation is
close to the level of the noise, the data hint at a slight correlation in the magnitudes
that could be the rotation of WeBo\,1's primary star.  While we cannot claim to have definitely detected this variability, the data suggest that
it is real.

If---as speculated by BPW03---the variability is the result of starspots on a
fast-rotating primary, the starspots may have changed over the course of
five years of observing, therefore muddying any periodicity in the full data set. But the variation would be 
marginally detectable over the short time span of the most recent observations.
Only more regular and systematic observations will be able to confirm WeBo\,1's variation as well as
any long-term changes.

\section{Conclusions}
\label{s:conc}

Using archival \Swift\slash UVOT images of the planetary nebula WeBo\,1, we have
detected a surplus of ultraviolet flux that cannot be explained by any
single-star system.  Using spectral models, we estimate that the system consists
of a K giant (previously identified by BPW03) and a hot compact object with a
temperature of at least 40,000 K and a radius of less than 0.056 $R_{\sun}$. This
temperature is significantly lower than expected for a young planetary nebula,
which is likely due to inadequacies in the foreground dust model.  The spectral
models provide an excellent reproduction of the photometry of PG~1159$-$035, a
hydrogen-deficient pre-WD with a similar extreme temperature but minimal
reddening.  Modifying the assumed extinction curve to a lower
$R_V$ and a stronger blue bump improves the fit to the photometry and increases
the derived temperature to well over 100,000~K\null. UV spectroscopy and/or FUV
imaging would be needed to provide a firm constraint on the pre-WD
temperature (as well as the foreground extinction). Our results do, however,
confirm the speculation of BPW03 that WeBo\,1 is a binary system, consisting of
a cool barium star and a hot pre-WD companion.

We detect some variability in the optical passbands which would be consistent
with the variability reported by BPW03. This variability is most clearly shown
in the subset of the data during which WeBo\,1 was more regularly monitored.  The long-term
data are more ambiguous.  If this result is confirmed, it would rule out 
regular pulsations, companion heating or ellipsoidal
variation as explanations for the variability seen by BPW03.  Only changing starspots
would produce a clear signal over a short period of time but a blurred signal over a longer
epoch.  However, the
measured variation of WeBo\,1 is comparable to the precision afforded by our
photometry. More thorough and precise photometric monitoring is needed to
determine if WeBo\,1's putative variation is the result, as speculated by BPW03,
of starspots on a rapidly rotating primary.

Finally, it is worth noting that all of the data in this study were obtained
serendipitously during the study of other objects.  This hints at the promise
held by the ongoing UV sky survey for discovering and characterizing
serendipitous sources.

\acknowledgements

The authors acknowledge sponsorship at PSU by NASA contract NAS5-00136.

\bibliographystyle{apj}

\end{document}